\documentclass{reducedws} 

\begin{document}

\title{Jet Tomography in Heavy Ion Collisions}

\author{Urs Achim Wiedemann}

\address{CERN TH Division, CH-1211 Geneva 23\\
E-mail: Urs.Wiedemann@cern.ch}

\maketitle

\abstracts{
We review recent calculations of the probability 
that a hard parton radiates an additional energy
fraction $\Delta E$ due to scattering in spatially extended matter,
and we discuss their application to the suppression of 
leading hadron spectra in heavy ion collisions at collider energies.}

\section{Introduction} 
\label{sec1}
Experiments at the Relativistic Heavy Ion Collider RHIC in Brookhaven
show for the first time a significant quenching of high-$p_\perp$
leading hadron spectra. In particular, transverse momentum spectra  for
neutral pion \cite{Adcox:2001jp,Mioduszewski:2002wt} and charged hadron 
\cite{Adcox:2001jp,Adler:2002xw} are suppressed 
if compared to spectra in p+p collisions rescaled by the number
of binary collisions. This suppression is most pronounced (up to 
a factor $\sim 5$) in central Au+Au collisions and smoothly 
approaches the binary scaling case with decreasing centrality.
The azimuthal anisotropy $v_2(p_\perp)$ of hadroproduction
stays close to maximal up to the highest transverse 
momentum \cite{Adler:2002ct}. Moreover, the disappearance
of back-to-back high-$p_\perp$ hadron correlations \cite{Adler:2002tq}
provides an additional indication that final state medium effects
play a decisive role in hadroproduction up to $p_\perp \sim 10$ GeV.

Parton energy loss 
has been proposed to account for the small nuclear modification 
factor \cite{Wang:2002ri}, the
azimuthal anisotropy \cite{Gyulassy:2000gk}
and the disappearance of dijets \cite{Muller:2002fa,Hirano:2003hq}.
Several studies (see references given in \cite{CarlosUrs})
indicate, however, that in the kinematical
regime relevant for RHIC ($p_\perp < 12$ GeV),  
$p_\perp$-broadening, shadowing, formation time 
and possibly other effects contribute significantly to the high-$p_\perp$
nuclear modification as well. 

Here, we discuss the status of parton energy loss calculations and
their comparison to data in the kinematical $p_\perp$-range probed 
at RHIC.

\section{Medium-induced gluon radiation from a static medium} 
\label{sec2}

Several groups
\cite{Baier:1996kr,Zakharov:1996fv,Wiedemann:2000za,Gyulassy:2000er} 
calculated recently the modification of the elementary splitting processes 
$q \to qg$ and $g \to gg$ due to multiple scattering. 
The inclusive energy distribution of gluon radiation off an in-medium 
produced parton takes the form
\cite{Wiedemann:2000za,Wiedemann:2000tf}
\begin{eqnarray}
  \omega\frac{dI}{d\omega}
  &=& {\alpha_s\,  C_R\over (2\pi)^2\, \omega^2}\,
    2{\rm Re} \int_{\xi_0}^{\infty}\hspace{-0.3cm} dy_l
  \int_{y_l}^{\infty} \hspace{-0.3cm} d\bar{y}_l\,
   \int d^2{\bf u}\,  \int_0^{\chi \omega}\, d^2{\bf k}\, 
  e^{-i{\bf k}_\perp\cdot{\bf u}}   \,
  e^{ -\frac{1}{2} \int_{\bar{y}_l}^{\infty} d\xi\, n(\xi)\,
    \sigma({\bf u}) }\,
  \nonumber \\
  && \times {\partial \over \partial {\bf y}}\cdot
  {\partial \over \partial {\bf u}}\,
  \int_{{\bf y}=0}^{{\bf u}={\bf r}(\bar{y}_l)}
   \hspace{-0.5cm} 
   {D}{\bf r}
   \exp\left[ i \int_{y_l}^{\bar{y}_l} \hspace{-0.2cm} d\xi
        \frac{\omega}{2} \left(\dot{\bf r}^2
          - \frac{n(\xi) \sigma\left({\bf r}\right)}{i\, \omega} \right)
                      \right]\, ,
    \label{eq1}
\end{eqnarray}
where gluons of transverse momentum $k_\perp < \chi\, \omega$ are
included. The radiation of hard quarks or gluons differs by 
the Casimir factor $C_R = C_F$ or $C_A$, respectively.  The 
properties of the medium enter eq. (\ref{eq1}) by the product of
the time-dependent density $n(\xi)$ of scattering centers times 
the strength of a single elastic scattering 
$\sigma({\bf r}) = 2 \int \frac{d{\bf q}}{(2\pi)^2}\,
                    \vert a({\bf q})\vert^2\, 
                    \left(1 - e^{i{\bf q}\cdot {\bf r}}\right)$, 
where  $\vert a({\bf q})\vert^2$ denotes the elastic high-energy 
cross section of a single scatterer.

For explicit calculations, one has to approximate the path integral
in (\ref{eq1}). This is done either by a saddle point approximation
\cite{Zakharov:1996fv} obtained in the 
medium-induced \underline{soft multiple scattering approximation}, 
\begin{eqnarray}
  n(\xi)\, \sigma({\bf r}) \simeq \frac{1}{2}\, \hat{q}(\xi)\, {\bf r}^2\, .
  \label{eq2}
\end{eqnarray}
Here, the only medium-dependent quantity is the transport 
coefficient \cite{Baier:1996sk} $\hat{q}(\xi)$ which
characterizes the transverse momentum squared $\mu^2$ transferred to 
the projectile per mean free path $\lambda$.
Alternatively, one can proceed in the \underline{opacity expansion}
which amounts to expanding the path integral in (\ref{eq1}) in powers
of the elastic scattering center
\cite{Wiedemann:2000za,Gyulassy:2000er}
\begin{eqnarray}
 &&\hspace{-0.3cm}K({\bf r},y_l;{\bf \bar r},\bar{y}_l) =
   {D}{\bf r}
   \exp\left[ i \int_{y_l}^{\bar{y}_l} \hspace{-0.2cm} d\xi
        \frac{\omega}{2} \left(\dot{\bf r}^2
          - \frac{n(\xi) \sigma\left({\bf r}\right)}{i\, \omega} \right)
                      \right]
 \label{eq3} \\
 &&\hspace{-0.3cm} \quad =
    K_0({\bf r},y_l;{\bf \bar r},\bar{y}_l)
 - \int\limits_{z}^{z'} {\it d}\xi\, n(\xi) 
 \int {\it d}\rho\,
 K_0({\bf r},y_l;\rho,\xi) \frac{\sigma(\rho)}{2}\, 
   K_0(\rho,\xi;{\bf \bar r},\bar{y}_l) + \dots\, .
 \nonumber 
\end{eqnarray}
%
\begin{figure*}[h]
\hspace*{-0.4cm}\includegraphics[width=62mm]{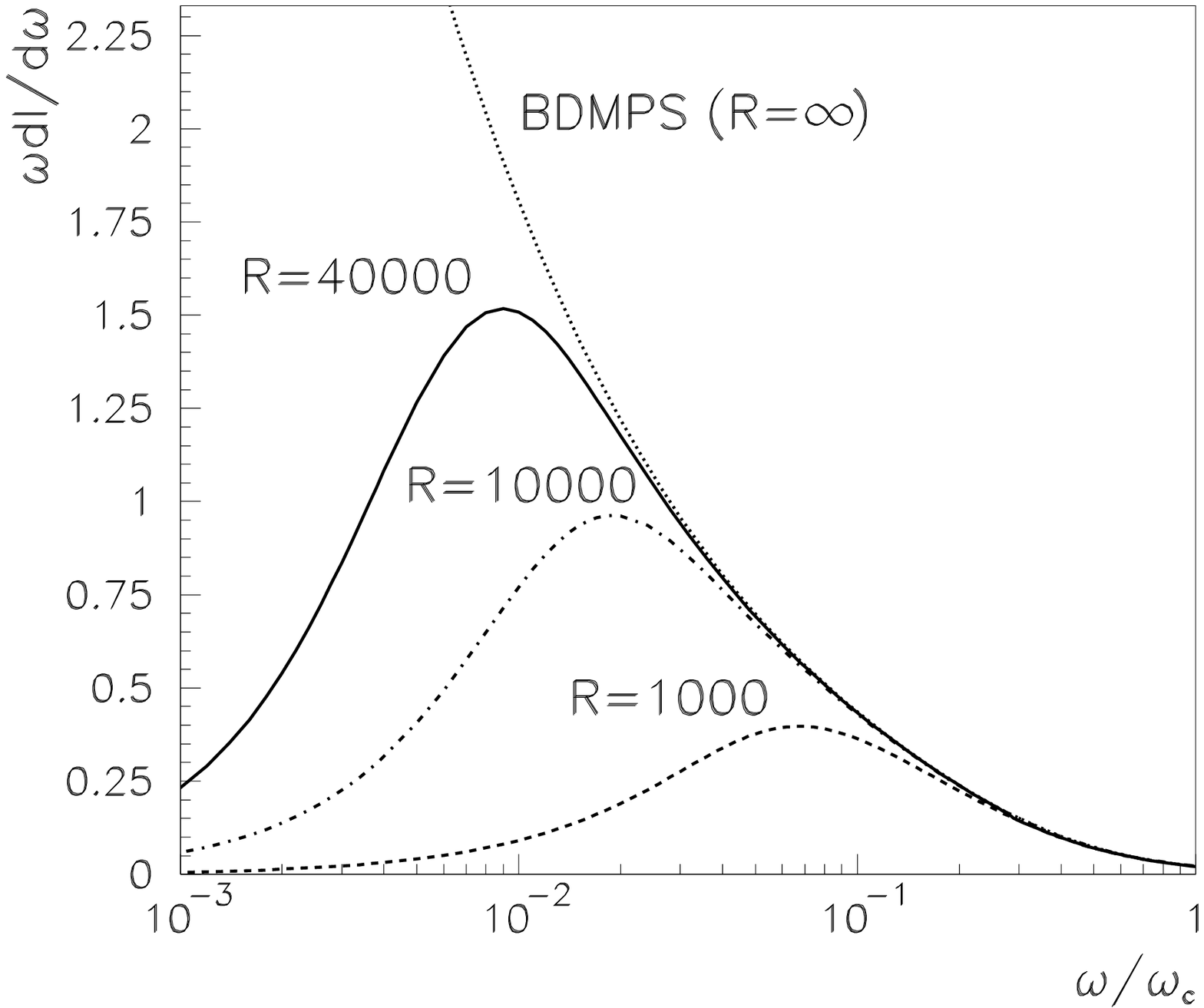}
\hspace*{-0.2cm}\includegraphics[width=56mm]{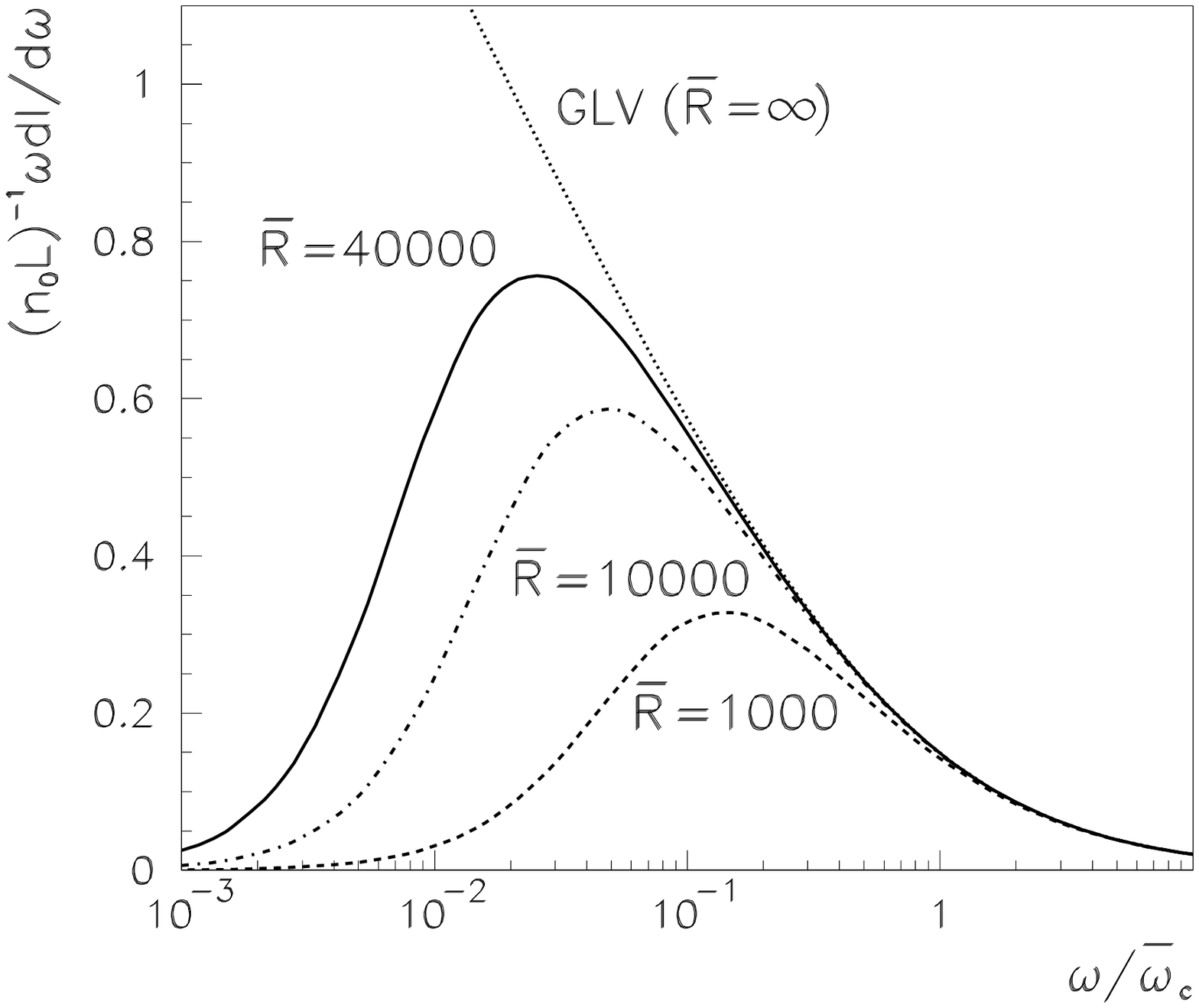}
\caption{The gluon energy distribution in the multiple soft (LHS) and
single hard (RHS) scattering approximation calculated for different
values of the kinematical constraints $R$, $\bar{R}$ respectively.
Fig. from Ref.\protect\cite{CarlosUrs}.}
\label{fig1}
\end{figure*}

Fig.\ref{fig1} compares numerical results for the gluon energy distribution
obtained in these two approximations. Qualitatively, this figure can
be understood by estimating to what degree initial state gluons 
{\it decohere} from the partonic projectile. Decoherence depends
on the relative phase $\varphi$ accumulated by the gluon due to
scattering. This phase grows with the transverse momentum
accumulated by the emitted gluon.

\noindent
\underline{For multiple soft scattering}, we have \cite{Baier:2001yt}
\begin{equation}
  \varphi = \Bigg\langle \frac{k_\perp^2}{2\omega}\, \Delta z \Bigg\rangle
  \sim \frac{\hat{q}\, L}{2\omega} L = \frac{\omega_c}{\omega}\, ,
  \label{eq4}
\end{equation}
which defines the ``characteristic gluon frequency'' 
$\omega_c = \frac{1}{2}\, \hat{q}\, L^2$. Given the number
$N_{\rm coh}$ of scattering centers which 
add coherently in the gluon phase (\ref{eq4}), one find 
$k_\perp^2 \simeq N_{\rm coh}\, \mu^2$. With
the coherence time of the emitted gluon, 
$t_{\rm coh} \simeq \frac{\omega}{k_\perp^2} \simeq 
\sqrt{\frac{\omega}{\hat{q}}}$
and $N_{\rm coh} = \frac{t_{\rm coh}}{\lambda} = 
\sqrt{\frac{\omega}{\mu^2\, \lambda}}$, one estimates for the
gluon energy spectrum per unit pathlength
\begin{equation}
  \omega \frac{dI^{(mult)}}{d\omega\, dz} \simeq 
  \frac{1}{N_{\rm coh}}\, 
  \omega \frac{dI^{\rm 1\, scatt}}{d\omega\, dz} \simeq
  \frac{\alpha_s}{t_{\rm coh}}
  \simeq \alpha_s\, \sqrt{\frac{\hat{q}}{\omega}}\, .
  \label{eq5}
\end{equation}
This $1/\sqrt{\omega}$-energy dependence agrees with the 
small-$\omega$ behaviour in Fig.\ref{fig1}. It is cut off for
$\omega > \omega_c$ where the phase (\ref{eq4}) is smaller than
unity and the reduced decoherence suppresses gluon emission.

\noindent
\underline{For single hard scattering} ($N=1$ opacity expansion)
with momentum transfer $\mu$, decoherence occurs if the typical 
gluon formation time $\bar{t}_{\rm coh} = \frac{2\omega}{\mu^2}$ 
is smaller than the typical distance $L$ between the production 
point of the parton and the position of the scatterer. The relevant 
phase is 
\begin{equation}
  \gamma = \frac{L}{\bar{t}_{\rm coh}} \equiv \frac{\bar{\omega}_c}{\omega}
  \, ,
  \label{eq6}
\end{equation}
which defines the characteristic gluon energy 
$\bar\omega_c = \frac{1}{2} \mu^2\, L$.
The gluon energy spectrum per unit pathlength can be estimated
in terms of the coherence time $\bar{t}_{\rm coh}$,
\begin{equation}
  \omega \frac{dI^{N=1}}{d\omega\, dz} \simeq 
  \frac{\alpha_s}{\bar{t}_{\rm coh}}
  \simeq \alpha_s\, \frac{\mu^2}{\omega}\, .
  \label{eq7}
\end{equation}
The full calculation in Fig.\ref{fig1}
agrees with this $1/\omega$-dependence in the range 
$\omega > \bar{\omega}_c$. 

In QCD, collinear gluons are hard and softer gluons tend to
be emitted under larger angles. As a consequence, the limitations
on transverse momentum phase space translate into a depletion of 
the infrared region of the gluon energy distribution, seen in 
Fig.\ref{fig1}. The estimates (\ref{eq5}) and (\ref{eq7}) determine
the dependence on the characteristic gluon energies $\omega_c$
($\bar{\omega}_c$) only. They do not include the constraint on
transverse phase space. The latter is determined by the
parameters \cite{Salgado:2002cd}
\begin{equation}
  R_{\chi} = \frac{1}{2}\, \hat{q}\, \chi^2\, L^3
  = \chi^2\, \omega_c\, L\, ,\qquad
  \bar{R}_{\chi} = \frac{1}{2}\, \chi^2\, \mu^2\, L^2
  = \chi^2\, \bar{\omega}_c\, L\, .
  \label{eq8}
\end{equation}
Clearly, keeping $\omega_c$ ($\bar{\omega}_c$) fixed and taking
$R$ ($\bar{R}$) to infinity amounts to the limit of infinite
pathlength. In this limit, the projectile can accumulate an 
arbitrarily large medium-induced transverse momentum and the
limit on transverse momentum phase space is removed. For a
medium of finite size, however, gluon emission at angles
$\Theta_c^2 \simeq \frac{\langle k_\perp^2 \rangle_{\rm med}}
{\omega^2}$ is suppressed since the emitted gluons
are sensitive to the kinematical constraint 
$k_\perp \leq O(\omega)$. In the multiple soft scattering
approximation, this translates into
\begin{equation}
  \Theta_c^2 
  \simeq \frac{\sqrt{\omega \hat{q}}}{\omega^2}
  \simeq \left( \frac{\omega}{\omega_c}\right)^{-3/2}
  \frac{1}{R} \sim 1\, \Longrightarrow
    \frac{\hat{\omega}}{\omega_c} \propto 
        \left(\frac{1}{R} \right)^{2/3}\, .
  \label{eq9}
\end{equation}
The position of the maximum of $\omega \frac{dI^{(mult)}}{d\omega}$ 
as a function of $R$ is consistent with this dependence
on $\hat{\omega}$, see Fig.~\ref{fig1}.

For the single hard scattering approximation, the 
corresponding estimate for the infrared cut-off $\hat{\omega}$
due to transverse momentum phase space constraints reads
\begin{equation}
  \Theta_c^2 \simeq \frac{\mu^2}{\hat{\omega}^2}
  \simeq \left(\frac{\bar\omega_c}{\hat{\omega}} \right)^2 
         \frac{1}{\bar{R}} \sim 1\quad \Longrightarrow
         \quad \frac{\hat{\omega}}{\bar\omega_c} 
               \propto \frac{1}{\sqrt{\bar{R}}}  \, .
  \label{eq10}
\end{equation}
The position of the maximum of $\omega \frac{dI^{N=1}}{d\omega}$
in Fig.~\ref{fig1} changes $ \propto \frac{1}{\sqrt{\bar{R}}}$, in
accordance with this estimate. We thus have a semi-quantitative 
understanding of how phase space constraints deplete the non-perturbative
soft region of the medium-induced gluon energy distribution.
This suppression of the non-perturbative small-$\omega$ contributions 
helps to make the calculation of medium-induced energy loss 
perturbatively stable.

In Ref.\cite{CarlosUrs}, we compare the single hard and multiple soft
scattering approximations of (\ref{eq1}) in detail. In
general, one finds that the gluon energy distribution is significantly
harder in the single hard scattering approximation. For example,
the average energy loss
$\Delta E = \int d\omega\, \omega\, \frac{dI}{d\omega}$
receives in the multiple soft scattering approximation 
a dominant contribution from the region $\omega < \omega_c$. 
In contrast, the dominant contribution comes from the hard
region $\omega > \bar{\omega}_c$ in the single hard scattering 
approximation.

Remarkably, although both approximations emphasize different
kinematical regions, the observable results are quantitatively
comparable if comparable sets of model parameters are used. 
To relate the model parameters in both approximations, one
observes that up to logarithmic accuracy
$\mu^2\, n_0\, L \simeq \hat{q}\, L$, where $n_0L$ determines the
average number of scattering centers within the in-medium
pathlength $L$. This implies
\begin{equation}
  R \simeq (n_0L)\, \bar{R}\, ,\qquad 
  \omega_c \simeq (n_0L)\, \bar{\omega}_c\, .
  \label{eq11}
\end{equation}
In the multiple soft scattering approximation, one uses the
transport coefficient $\hat{q}$ to characterize the average transverse 
momentum squared per unit pathlength. In the opacity expansion, one
specifies not only this information, but additionally
the average number of scattering centers in which 
this momentum transfer takes place. As a consequence, the opacity
expansion has one additional model parameter $n_0L$. Numerically,
we find that for the choice $n_0L = 3$ in (\ref{eq11}), the energy 
density distribution, quenching weight and even the angular dependence 
of the average energy loss are quantitatively comparable \cite{CarlosUrs}. 
We found deviations from this relation 
for those quantities for which due to kinematical constraints
only the region $\omega < \omega_c$ contributes in {\it both} 
approximations, see below. This is in particular the case for
the comparison to RHIC data in Fig.\ref{fig4}, where 
$\omega < E < \omega_c$.  

\section{Quenching Weights}
\label{sec3}

If gluons are emitted independently by a hard parton, 
then the probability $P(\Delta E)$ that this parton loses an
additional energy fraction $\Delta E$ can be calcualted from
the normalized sum of the emission probabilities
for an arbitrary number of $n$ gluons which carry away a total
energy $\Delta E$:
\begin{eqnarray}
  P(\Delta E) = \sum_{n=0}^\infty \frac{1}{n!}
  \left[ \prod_{i=1}^n \int d\omega_i \frac{dI(\omega_i)}{d\omega}
    \right]
    \delta\left(\Delta E - \sum_{i=1}^n \omega_i\right)
    \exp\left[ - \int d\omega \frac{dI}{d\omega}\right]\, .
   \label{eq12}
\end{eqnarray}
The summation over arbitrarily many gluon emissions in 
(\ref{eq12}) can be performed analytically by Laplace 
transformation~\cite{Baier:2001yt,Salgado:2002cd,Arleo:2002kh}. 
In general, one finds
a discrete and a continuous part,
\begin{equation}
  P(\Delta E) = p_0\, \delta(\Delta E) + p(\Delta E)\, .
   \label{eq13}
\end{equation}
The discrete weight $p_0$ may be viewed as the probability that 
no additional gluon is emitted due to in-medium scattering 
and hence no medium-induced energy loss occurs. 
For finite in-medium pathlength, there is always a finite 
probability $p_0$ that the projectile is not affected by the 
medium. For infinite in-medium pathlength, however, one finds 
$ \lim_{R\to\infty} p_0 = 0$.
\vspace{-1.0cm}
\begin{figure}[h]\epsfxsize=12.0cm
\centerline{\epsfbox{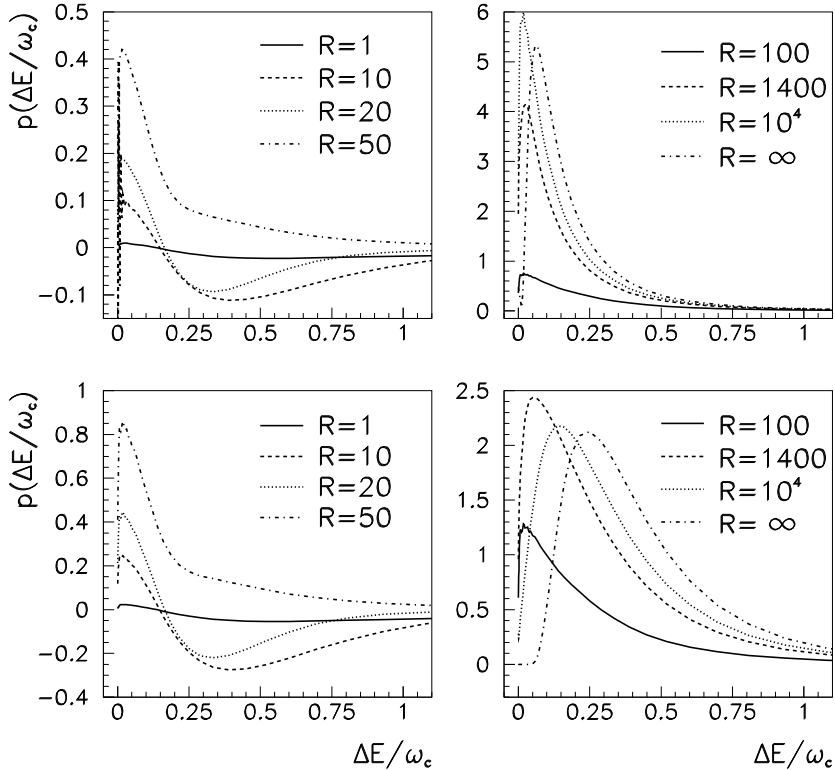}}
\caption{The continuous part of the quenching weight 
(\protect\ref{eq13}), calculated in the multiple soft scattering limit
for a hard quark (upper row) or hard gluon (lower row).
Fig. from Ref.\protect\cite{CarlosUrs}.
}\label{fig2}
\end{figure}

In Fig. \ref{fig2}, we show the continuous part of the quenching
weight (\ref{eq13}) in the multiple soft scattering approximation.
The parton energy loss $\Delta E$ is generally seen to increase 
with increasing characteristic gluon energy $\omega_c$ or increasing 
factor $R$. This implies that $\Delta E$ grows with increasing 
momentum transfer $\hat{q}$ from the medium, and with increasing
in-medium pathlength $L$, as naively expected.  

To understand the negative contributions of the quenching weight
for small values $R$, one recalls that
the gluon energy distribution $\omega \frac{dI}{d\omega}$ 
calculated in (\ref{eq1}) is only the medium-induced modification of 
a radiation pattern $\frac{dI^{(vac)}}{d\omega}$ which occurs in the
absence of a medium,
\begin{equation}
 \omega \frac{dI^{(tot)}}{d\omega} = 
 \omega \frac{dI^{(vac)}}{d\omega} +
 \omega \frac{dI}{d\omega}\, .
 \label{eq14}
\end{equation}
By writing the corresponding probabilities (\ref{eq11}) in Mellin
space, one shows that
\begin{equation}
  P^{(tot)}(\Delta E) = \int_0^\infty  d\bar{E} \, 
  P(\Delta E - \bar{E}) \, 
   P^{(vac)}(\bar{E})\, . 
   \label{eq15}
\end{equation} 
The probability $P^{(tot)}(\Delta E)$ is normalized to unity and it is 
positive definite. In contrast, the medium-induced modification of this 
probability, $P(\Delta E)$, is a generalized probability. It can take
negative values for some range in $\Delta E$, as long as its normalization
is unity,
\begin{equation}
  \int_0^\infty  d\bar{E} \,  P(\bar{E})
  = p_0 + \int_0^\infty  d\bar{E} \,  p(\bar{E}) = 1\, .
  \label{eq16}
\end{equation}
The discrete weight $p_0$ is found to coincide with this normalization
condition. A qualitatively similar behaviour of the quenching weights
is found in the single hard scattering approximation.

\section{Angular dependence of radiation probability}
\label{sec4}
\begin{figure}[h]\epsfxsize=12.7cm
\centerline{\epsfbox{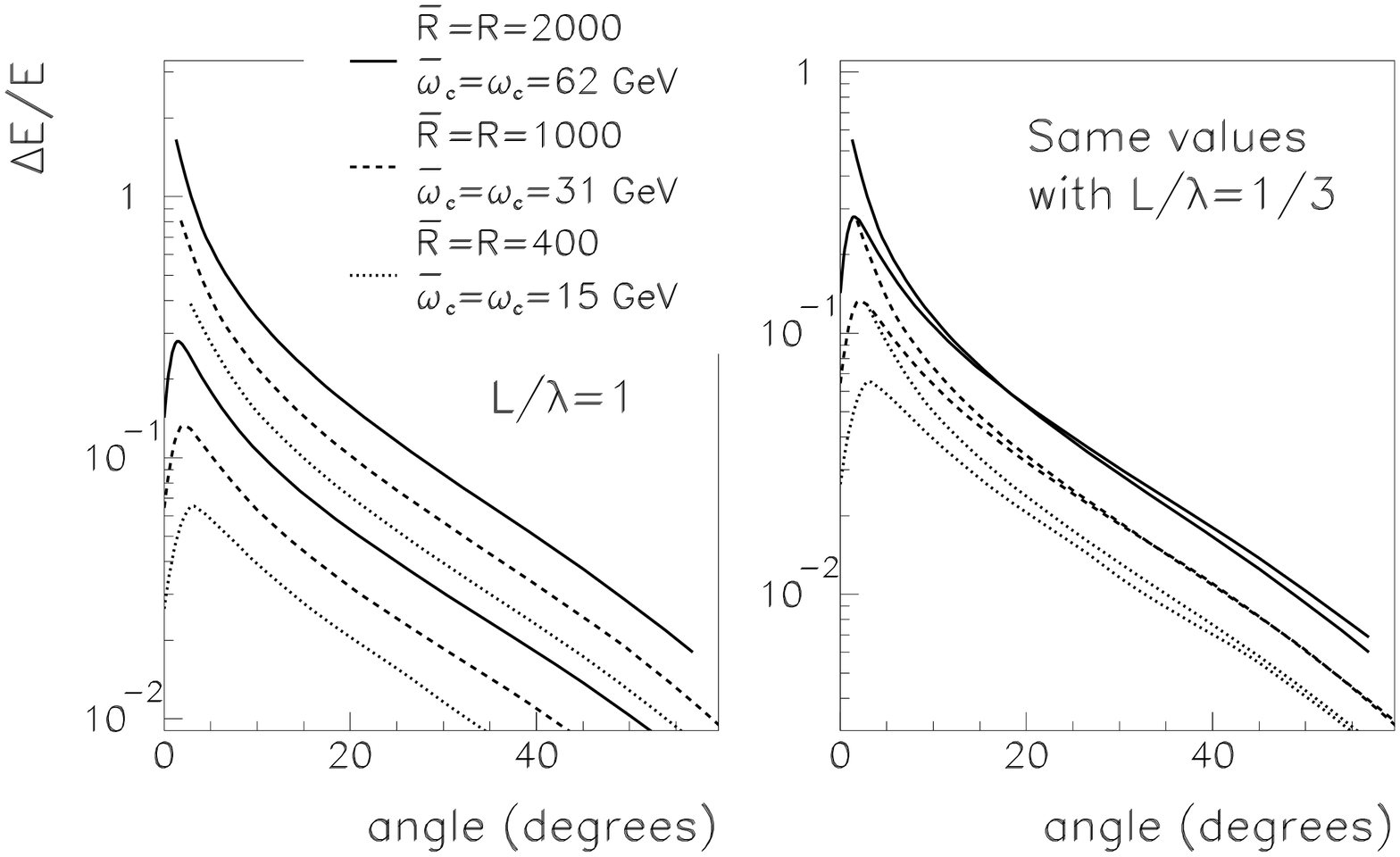}}
\vspace{0.5cm}
\caption{The average energy loss (\protect\ref{eq17}) radiated
outside an angle $\Theta$ as calculated in the multiple soft
(lower three lines) and single hard (upper three lines) 
scattering approximation. Fig. from Ref.\protect\cite{CarlosUrs}.
}\label{fig3}
\end{figure}

The quenching weights discussed above allow to
calculate the average energy loss {\it outside} 
an opening angle $\Theta$,
\begin{eqnarray}
  \langle \Delta E \rangle(\Theta) 
  &=& \int d\omega\, \omega\, 
   \frac{dI^{>\Theta}}{d\omega}(\omega_c,R=\omega_cL)
   \nonumber \\
  &=& \int d\bar{E}\, \bar{E}\, 
  \left[ P(\bar{E},\omega_c,R=\omega_cL) 
         - P(\bar{E},\omega_c,R_\chi=\chi^2\omega_cL) 
         \right]\, .
  \label{eq17}
\end{eqnarray}
This calculation is straightforward since a finite emission
angle $\Theta$ can be taken into account simply by reducing
the kinematical constraint $k_\perp < \omega$ in (\ref{eq1})
to $k_\perp < \chi\, \omega$, $\chi = \sin\Theta$. 
In Fig.\ref{fig3}, we compare the resulting angular radiation
patterns obtained in the single hard and multiple soft
scattering approximation. In agreement with the discussion
following eq. (\ref{eq11}) above, we find that for comparable 
sets of model parameters $\omega_c$, $R$ and $\bar{\omega}_c$, 
$\bar{R}$, $n_0\, L$ respectively, the multiple soft and single hard 
scattering approximations lead to a comparable angular 
dependence of $\langle \Delta E\rangle(\Theta)$ for $\Theta > 10^{\circ}$. 
For smaller angles, deviations persist which one can understand
quantitatively \cite{CarlosUrs} and which should be regarded as an
intrinsic uncertainty of this calculation. 
\vspace{-0.5cm}
\vspace{-0.5cm}
\section{Application}
\label{sec5}
The quenching weight $P(\Delta E)$ determines the quenching
factor \cite{Baier:2001yt}
\begin{eqnarray}
 Q(p_\perp)&=&
\int d{\Delta E}\, P(\Delta E)\left(
{d\sigma^{\rm vac}(p_\perp+\Delta E)/ dp^2_\perp}\over
{d\sigma^{\rm vac}(p_\perp)/ dp^2_\perp}\right)\, ,
 \label{eq18}
\end{eqnarray}
which determines the reduction of transverse momentum leading 
hadron spectra due to medium-induced energy loss. Alternatively,
the quenching weight can be used to determine medium-modified
fragmentation functions~\cite{Wang:1996yh}
\begin{eqnarray}
  D_{h/q}^{(\rm med)}(x,Q^2) 
  = \int_0^1 d\epsilon\, P(\epsilon)\,
  \frac{1}{1-\epsilon}\, 
  D_{h/q}(\frac{x}{1-\epsilon},Q^2)\, .
  \label{eq19}
\end{eqnarray}
%
\begin{figure}[h]\epsfxsize=9.0cm
\centerline{\epsfbox{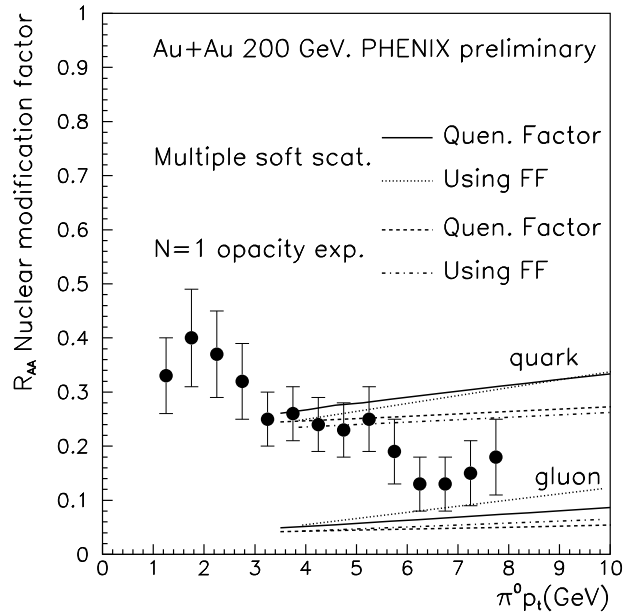}}
\caption{The nuclear modification factor for $\pi^0$-production
\protect\cite{Mioduszewski:2002wt}
compared to model calculations involving parton energy loss
only. Curves present the quenching factor (\protect\ref{eq18})
and the suppression factor (\protect\ref{eq20}) obtained from
medium-modified fragmentation functions. They are given in the
limiting cases where all parent partons are either quarks 
(upper lines) or gluons (lower lines). Calculations in
the multiple soft scattering approximation use $R = 2000$, 
$\omega_c = 67.5$ GeV, corresponding to $\hat{q} = 0.75
\frac{{\rm GeV}^2}{\rm fm}$ and $L=6$ fm. In the single
hard scattering approximation, we use $\bar{R} = R$, 
$\bar{\omega}_c = \omega_c$. Fig. from Ref. \protect\cite{CarlosUrs}.
}\label{fig4}
\end{figure}

From these, the reduction of leading transverse momentum
spectra can be calculated \cite{CarlosUrs} by convoluting 
with the perturbative
hard matrix elements which are approximately proportional to
$x^6$,
\begin{equation}
  R_{ff}(p_\perp)
  = \frac{ x_{\rm max}^6 D_{h/q}^{(\rm med)}(x_{\rm max},p_\perp^2)}{
                   x_{\rm max}^6 D_{h/q}(x_{\rm max},p_\perp^2)}
  \Bigg\vert_{p_\perp = x_{\rm max}\, E_q}\, .
  \label{eq20}
\end{equation}

In Fig.\ref{fig4}, we compare the data on the suppression of the
$\pi^0$-spectra to suppression factors calculated from (\ref{eq18})
and (\ref{eq20}). We find that both definitions of the suppression
factors lead to quantitatively comparable results. Moreover, 
the single hard and multiple soft scattering approximation lead
to quantitatively comparable results for suitable choices of the
model parameters. Remarkably, we find that in the presence of
kinematical constraints depleting the infrared region of
$\omega \frac{dI}{d\omega}$ in Fig.\ref{fig1}, the 
$p_\perp$-dependence of the quenching factor flattens
considerably. This feature is necessary to find in Fig.\ref{fig4}
a shallow $p_\perp$-dependence which is consistent with the data.

The application of the current calculations of parton energy loss to
data below $p_\perp < 10$ GeV entails significant theoretical 
uncertainties. On the one hand, there are plausible competing
physics effects which may be relevant at $p_\perp < 10$ GeV
(see Ref.\cite{CarlosUrs} for further discussion). On the other hand,
the high-energy ``eikonal'' approximation used in the derivation
of (\ref{eq1}) becomes questionable if $\Delta E \sim E$
which is the case in Fig.\ref{fig4}. This should motivate 
improved calculations of (\ref{eq1}) for which the finite
energy effects of the partonic projectile are taken into
account.


\end{document}